\documentclass[pdflatex]{moriond}
\usepackage{amssymb,amsmath,epsfig}

\def\Journal#1#2#3#4{{#1} {\bf #2}, #3 (#4)}

\def\PRL{\em Phys. Rev. Lett.}
\def\PRD{{\em Phys. Rev.} D}

\def\be{\begin{equation}}
\def\ee{\end{equation}}
\def\bea{\begin{eqnarray}}
\def\eea{\end{eqnarray}}

\DeclareMathOperator{\tr}{Tr}
\newcommand{\sT}{{\scriptscriptstyle T}}
\newcommand{\Photo}{}
\renewcommand{\d}{\mathrm{d}}
\newcommand{\bm}[1]{{\bf #1}}
\newcommand{\qT}{\bm{q}_\sT}
\newcommand{\mc}[1]{\mathcal{#1}}

\newcommand{\ce}[1]{{Eq.~\ref{#1}}}

\begin{document}
\vspace*{4cm}
\title{PROBING THE TRANSVERSE DYNAMICS AND POLARIZATION OF GLUONS INSIDE THE PROTON AT THE LHC}

\author{CRISTIAN PISANO}

\address{Nikhef and Department of Physics and Astronomy, VU University Amsterdam, De Boelelaan 1081, NL-1081 HV Amsterdam, The Netherlands}

\maketitle\abstracts{Transverse momentum dependent gluon distributions encode fundamental information on the structure of the proton. Here we show how they can be accessed in heavy quarkonium production in proton-proton collisions at the LHC. In particular, their first determination could come from the study of an isolated $J/\psi$ or $\Upsilon$ particle, produced back to back with a photon.}

\section{Formalism}
Transverse momentum dependent (TMD) gluon distributions inside an unpolarized proton are defined by the hadron matrix element of a correlator of the gluon
field strengths $F^{\mu\rho}(0)$ and $F^{\nu\sigma}(\xi)$. Expanding the gluon 
four-momentum as $p = x\,P + p_\sT + p^- n$, with $n$ being a lightlike vector conjugate to the momentum of the proton $P$, such correlator can be written as~\cite{Mulders:2000sh}  
\begin{eqnarray}
\label{GluonCorr}
\Phi_g^{\mu\nu}(x,\bm p_\sT )
& = &  \frac{n_\rho\,n_\sigma}{(p{\cdot}n)^2}
{\int}\frac{d(\xi{\cdot}P)\,d^2\xi_\sT}{(2\pi)^3}\
e^{ip\cdot\xi}\,
\langle P|\,\tr\big[\,F^{\mu\rho}(0)\,
F^{\nu\sigma}(\xi)\,\big]
\,|P \rangle\,\big\rfloor_{\xi\cdot n=0} \nonumber \\
& = &
-\frac{1}{2x}\,\bigg \{g_\sT^{\mu\nu}\,f_1^g (x,\bm p_\sT^2)
-\bigg(\frac{p_\sT^\mu p_\sT^\nu}{M_p^2}\,
{+}\,g_\sT^{\mu\nu}\frac{\bm p_\sT^2}{2M_p^2}\bigg)
\;h_1^{\perp\,g} (x,\bm p_\sT^2) \bigg \} ,
\end{eqnarray}
where gauge links have been omitted. The transverse projector  $g^{\mu\nu}$ is defined as $g^{\mu\nu}_{\sT} = g^{\mu\nu} - P^{\mu}n^{\nu}/P{\cdot}n-n^{\mu}P^{\nu}/P{\cdot}n$. Moreover, $p_{\sT}^2 = -\bm p_{\sT}^2$ and $M_p$ is the proton mass. The gluon correlator of an unpolarized proton can therefore be 
expressed in terms of two independent TMD distribution functions:  $f_1^g(x,\bm{p}_\sT^2)$ is the unpolarized one, while  $h_1^{\perp\,g}(x,\bm{p}_\sT^2)$ denotes the $T$-even, helicity-flip distribution of linearly polarized gluons, which satisfies the model-independent positivity bound~\cite{Mulders:2000sh}
\begin{equation}
\frac{\bm p_\sT^2}{2M_p^2}\,|h_1^{\perp g}(x,\bm p_\sT^2)|\le f_1^g(x,\bm p_\sT^2)\,.\label{eq:Bound}
\end{equation}
Like any TMD distribution, $h_1^{\perp\,g}$ might receive contributions from initial and final state interactions that can render it nonuniversal and even hamper its extraction in processes for which TMD factorization does not apply.  

\section{Phenomenology}

Several processes have been suggested to measure the experimentally unkown distributions $f_1^g$ and  $h_1^{\perp\,g}$. Although it has been discussed how to isolate the contribution from $h_1^{\perp \, g}$ by means of an azimuthal angular dependent weighting of the cross section for dijet production in hadronic collisions~\cite{Boer:2009nc}, TMD factorization is expected to be broken in this case due to the presence of both initial and final state interactions~\cite{Rogers:2010dm}. A theoretically cleaner and safer way would be to study 
dijet or heavy quark pair production in electron-proton collisions, for instance at a future Electron-Ion Collider~\cite{ep2jetLO1,ep2jetLO2}. Another process where the problem of factorization breaking is absent is $pp \to \gamma \gamma X$~\cite{Qiu:2011ai}, which however suffers from a huge background from $\pi^0$ decays and contaminations from quark-induced channels.

In the following we show how TMD gluon distributions can be probed in heavy quarkonium production at the LHC. TMD factorization should hold in this case, provided that the two quarks that form the bound state are produced in a colorless state already at short distances.

\subsection{Transverse momentum distributions of $C=+$ quarkonia}

We consider first the process $p(P_A)+p(P_B)\to {\cal Q}(q)+X$, where 
${\cal Q}$ is a heavy quark-antiquark bound state with $C=+$, and the four-momenta of the particles are given between brackets. Assuming TMD factorization, the corresponding cross section can be written as
\begin{eqnarray}
\d\sigma
& = &\frac{1}{2 s}\,\frac{d^3 \bm q}{(2\pi)^3\,2 q^0} 
{\int} \d x_a \,\d x_b \,\d^2\bm p_{a\sT} \,\d^2\bm p_{b\sT}\,(2\pi)^4
\delta^4(p_a{+} p_b {-} q)
 \nonumber \\
&&\qquad \times
\,  \Phi_g^{\mu\nu}(x_a {,}\,\bm p_{a \sT})\, \Phi_g^{\rho\sigma}(x_b {,}\,\bm p_{b \sT})\overline{\sum_{\rm colors}}
 {\cal A}_{\mu\rho}\, {\cal A}_{\nu\sigma}^*\,(p_a, p_b; q)\,,
\label{CrossSec}
\end{eqnarray}
with $s = (P_A + P_B)^2$ being the total energy squared in the hadronic center-of-mass frame and ${\cal A}$ denoting the hard scattering amplitude of the dominant subprocess $g(p_a)\,+\,g(p_b)\,\to\, {\cal{Q}}(q)$. The 
amplitude ${\cal A}$ is evaluated at order $\alpha_s^2$ within the framework of the color-singlet model. Color octet contributions should be negligible, according to nonrelativistic QCD arguments~\cite{ppJPsiLO}. For small
transverse momentum, $\qT^2 \ll M^2_{\cal Q}$, with $M_{\cal Q}$ being the quarkonium mass, the resulting transverse momentum distributions for $\eta_Q$ and $\chi_{Q0,2}$ ($Q=c$, $b$)  are  given by
\begin{eqnarray}
\frac{1}{\sigma(\eta_Q)} \, \frac{\d\sigma (\eta_Q)}{\d y\, \d \bm q_\sT^2} & = &  
\frac{{\cal C} [f_1^g f_1^g]}{\int \d\bm q_\sT^2 \, {\cal C} [f_1^gf_1^g] } \,\left [1-R(\bm q_\sT^2)\right ] \, ,\nonumber \\
\frac{1}{\sigma(\chi_Q)} \, \frac{\d\sigma (\chi_{Q 0})}{\d y \,\d \bm q_\sT^2}  & = &  \frac{{\cal C} [f_1^g f_1^g]}{\int \d\bm q_\sT^2 \, {\cal C} [f_1^gf_1^g] }\, \left [1+R(\bm q_\sT^2)\right ] \, ,\nonumber \\
\frac{1}{\sigma(\chi_Q)} \, \frac{\d\sigma (\chi_{Q 2})}{\d y\, \d \bm q_\sT^2}  & = & \frac{{\cal C} [f_1^g f_1^g]}{\int \d\bm q_\sT^2 \, {\cal C} [f_1^gf_1^g] }\,,
\end{eqnarray}
where $\sigma = \int \d \qT^2\, \d \sigma$ and $y$ is the rapidity of the quarkonium along the direction of the incoming protons. Furthermore, $x_{a,b} =  {M_{\cal Q}}/{\sqrt{s}}\, e^{\pm y}$,
\begin{equation}
R (\bm q_{\sT}^2) = \frac{\mathcal{C}\left[w\, h_{1}^{\perp \,g}\, h_{1}^{\perp \,g} \right ]}{\mathcal{C}\left[f_{1}^{g}\, f_{1}^{g}\right]}\,,\qquad w= \frac{1}{2M^{4}} \, \left [ (\bm p_{a\sT}\cdot\bm p_{b\sT})^{2}-\frac{1}{2} \, \bm p_{a\sT}^{2}
\bm p_{b\sT}^{2} \right ]\,,
\end{equation}
and we have used the following definition of convolution of two TMD distributions $f$ and $g$,  
\begin{eqnarray}
\mathcal{C}[w\, f\, g] & \equiv & \int d^{2}\bm p_{a\sT}\int d^{2}\bm p_{b\sT}\,
\delta^{2}(\bm p_{a\sT}+\bm p_{b\sT}-\bm q_{\sT})\, w(\bm p_{a\sT},\bm p_{b\sT})\, f(x_{a},\bm p_{a\sT}^{2})\, g(x_{b},\bm p_{b\sT}^{2})\,.\label{eq:Conv}
\end{eqnarray} 

\begin{figure*}[t]
\centering
{\includegraphics[width=0.36\textwidth]{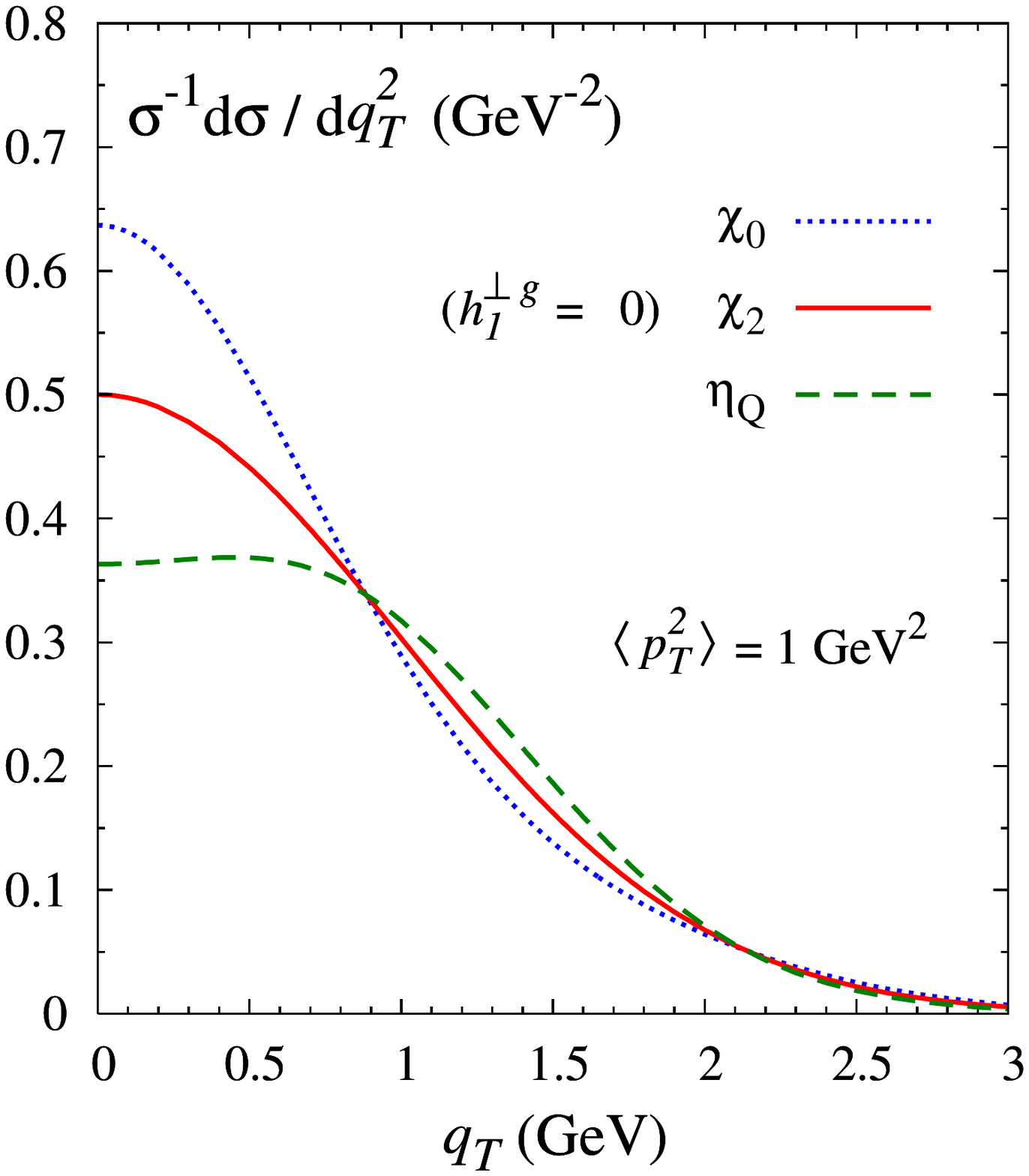}}
{\includegraphics[width=0.36\textwidth]{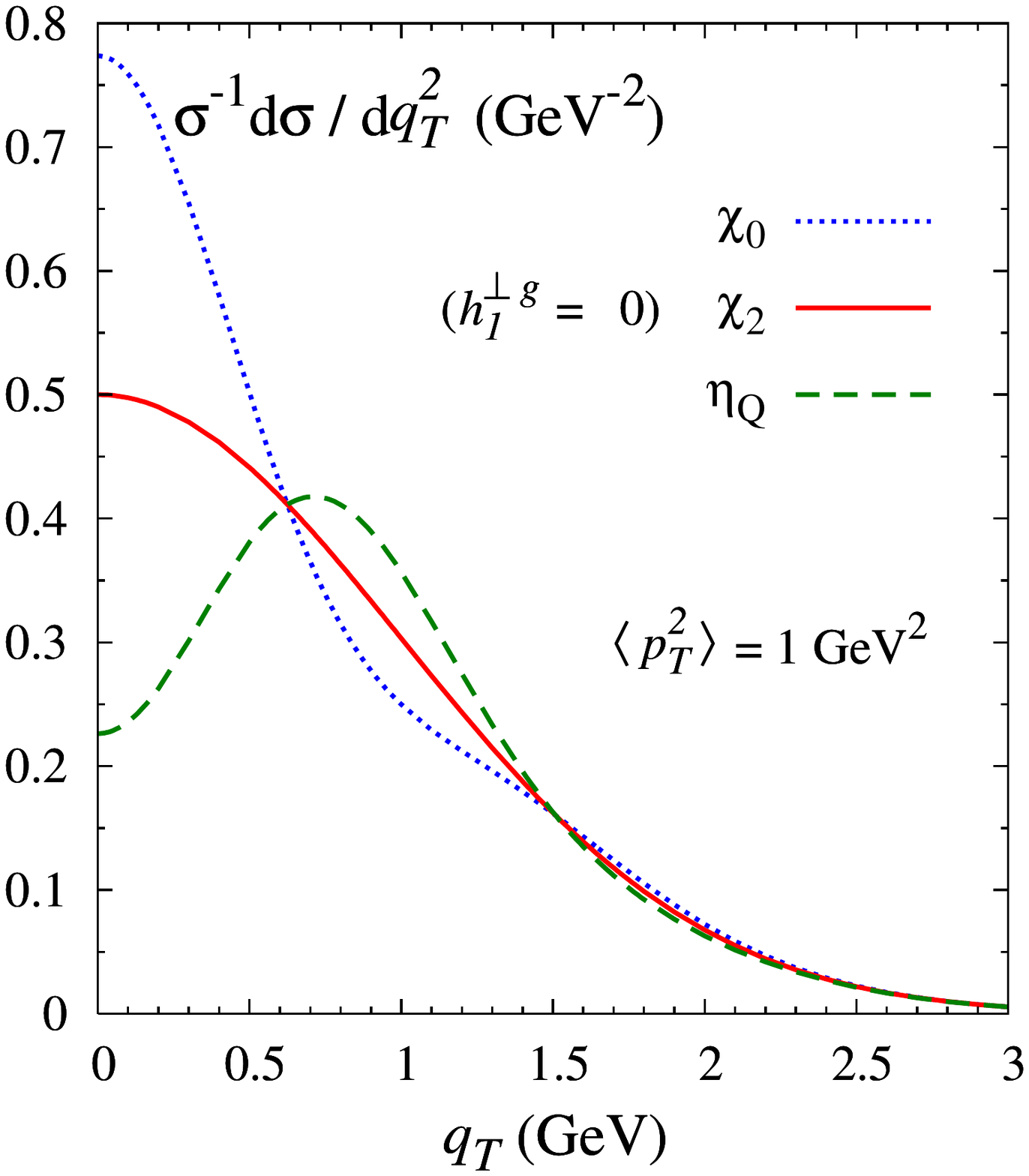}}
\hspace{0.5cm}
\caption{Transverse momentum distributions for $C=+$ quarkonia evaluated  at $y=0$, obtained using the input distributions in Eqs.~(\ref{eq:Gaussf1})-(\ref{eq:Gaussh1perp}), with $\langle p_\sT^2 \rangle = 1$ GeV$^2$ and two different values of $r$: $r=2/3$ (left) and $r=1/3$ (right).}
\label{fig:dsigmadqT}
\end{figure*}

Our numerical estimates are shown in Fig.\ref{fig:dsigmadqT}, where we have assumed that the gluon distributions have a simple Gaussian dependence on transverse momentum. Namely, 
\begin{equation}
f_1^g(x,\bm p_\sT^2) = \frac{f_1^g(x)}{\pi \langle  p_\sT^2 \rangle}\,
\exp\left(-\frac{\bm p_\sT^2}{\langle  p_\sT^2 \rangle}\right)\,,\label{eq:Gaussf1}
\end{equation}
where $f_1^g(x)$ is the collinear gluon distribution and the width
$\langle p_\sT^2 \rangle$ is taken to be independent  of $x$ and the energy scale, set by $M_{\cal Q}$. The bound in Eq.\ (\ref{eq:Bound}) is satisfied, although not everywhere saturated, by the form
\begin{equation}
h_1^{\perp g}(x,\bm p_\sT^2)=\frac{M_p^2 f_1^g(x)}{\pi \langle p_\sT^2 \rangle^2}\frac{2(1-r)}{r}\,
\exp\left(1 -\frac{1}{r}\,\frac{\bm p_\sT^2}{\langle p_\sT^2 \rangle}\right)\,,\label{eq:Gaussh1perp}
\end{equation}
with $0<r<1$. The distributions for $\eta_{c,b}$ and 
$\chi_{c,b\,0}$ are similar to the ones for a pseudoscalar and a scalar Higgs boson~\cite{TMDHiggs1,TMDHiggs2}, and can be used to extract $h_1^{\perp\,g}$, while $f_1^g$ can be accessed by looking at $\chi_{c,b\,2}$. A comparison among the different spectra could help to cancel out uncertainties. This experiment requires
forward detectors like the LHCb, which hopefully will be able to provide such data in the near future.

\subsection{$C=-$ quarkonium production in association with a photon}

Along the lines of the previous section, we study the process  $p(P_A)+p(P_B)\to {\cal Q}(P_{\cal Q}) + \gamma (P_{\gamma})+ X$, where now ${\cal Q}$ is a $C=-$ quarkonium ($J/\psi$ or $\Upsilon$) produced almost back to back with the photon. Hence the imbalance $\qT = \bm P_{{\cal Q}\sT}+P_{\gamma\sT}$ will be small, but not the individual transverse momenta of the two particles. No forward detector is therefore needed in this case. The  cross section has the following structure,
\begin{eqnarray}\label{eq:crosssection}
\frac{\d\sigma}{\d Q \d Y \d^2 \qT \d \Omega} 
  \propto 
  F_1\, \mc{C} \Big[f_1^gf_1^g\Big]+ F_3 \,\mc{C} \Big[w_3 f_1^g h_1^{\perp g} + x_a\! \leftrightarrow\! x_b \Big] \cos 2\phi + F_4  \mc{C} \left[w_4 h_1^{\perp g}h_1^{\perp g}\right]\cos 4\phi \,,
\end{eqnarray}
where  $Q$ and $Y$  are the invariant mass and the rapidity of the pair, to be measured, like $\bm q_{\sT}$, in the hadronic center-of-mass frame. On the other hand, the solid angle $\Omega=(\theta,\phi)$ is measured in the Collins-Soper frame, where the final pair is at rest and the $\hat x\hat z$-plane is spanned by $\bm P_A$ and $\bm P_B$, with the $\hat x$-axis set by their bisector. The transverse weights are given by 
\begin{equation}
 w_3 = \frac{\qT^2\bm p_{b\sT}^2 - 2 (\qT{\cdot}\bm p_{b \sT})^2}{2 M_p^2 \qT^2},
\qquad
 w_4 =  2\left[\frac{\bm p_{a\sT}{\cdot}\bm p_{b\sT}}{2M_p^2} - 
		\frac{(\bm p_{a\sT}{\cdot}\qT) (\bm p_{b\sT}{\cdot}\qT)}{M_p^2\qT^2}\right]^2 -\frac{\bm p_{a\sT}^2 \bm p_{b \sT}^2 }{4 M_p^4}\,,
\end{equation}
and the light-cone momentum fractions are $x_{a,b} =  \exp[\pm Y]\, Q/\sqrt{s}$.
Explicit expressions for $F_{1,3,4}$ can be found elsewhere~\cite{Qgamma}.
 We propose the measurement of the following three observables,
\begin{equation}
{\cal S}^{(n)}_{q_T} \equiv  \frac{\int \d\phi\, \cos(n\, \phi )\, \frac{\d\sigma}{\d Q \d Y \d^2 \qT \d \Omega}}
{\int \d \bm q_\sT^2 \int \d\phi \,\frac{\d\sigma}{\d Q \d Y \d^2 \qT \d \Omega}}\,,
\end{equation}
with  $n=0,2,4$, and where the $q_\sT^2$ integration in the denominator is 
up to $(Q/2)^2$. In this way we are able to single out the three terms in \ce{eq:crosssection}: 
\begin{align}\label{eq:qTdistrs}
{\cal S}^{(0)}_{q_T} =\frac{\mc{C}[f_1^g f_1^g]}
  {\int  \d \bm q_\sT^2\, \mc{C}[f_1^g f_1^g]},\quad
  {\cal S}^{(2)}_{q_T}= 
  \frac{F_3\, \mc{C}[w_3 f_1^g h_1^{\perp g} + x_a \leftrightarrow x_b]}
  {2 F_1 \int  \d \bm q_\sT^2\, \mc{C}[f_1^g f_1^g]},\quad 
{\cal S}^{(4)}_{q_T}\!=  
 \frac{F_4\, \mc{C}[w_4 h_1^{\perp g} h_1^{\perp g}]}
  {2 F_1 \int \d \bm q_\sT^2\, \mc{C}[f_1^g f_1^g]}\,.
\end{align}

\begin{figure*}[t]
\centering
{\includegraphics[width=0.31\textwidth]{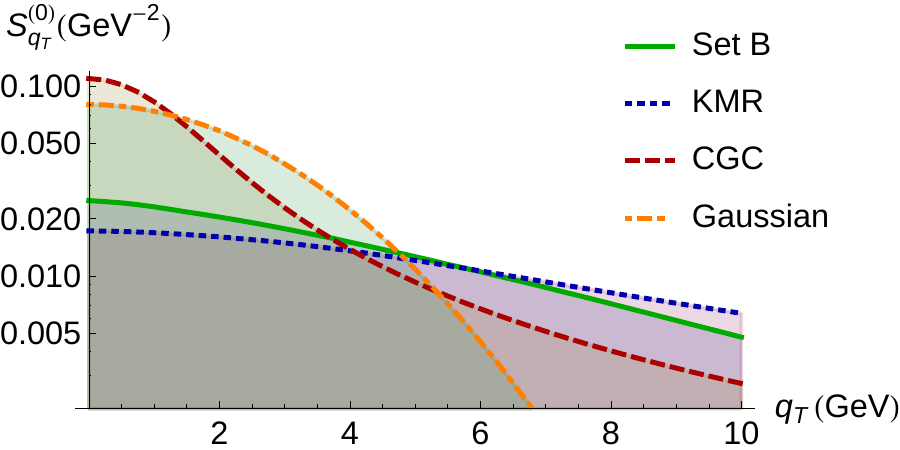}}
\hspace{0.3cm}
{\includegraphics[width=0.31\textwidth]{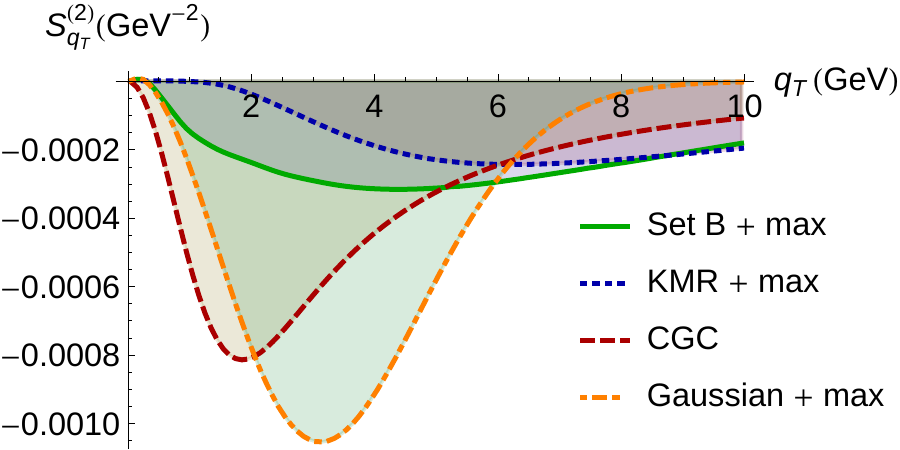}}
\hspace{0.3cm}
{\includegraphics[width=0.31\textwidth]{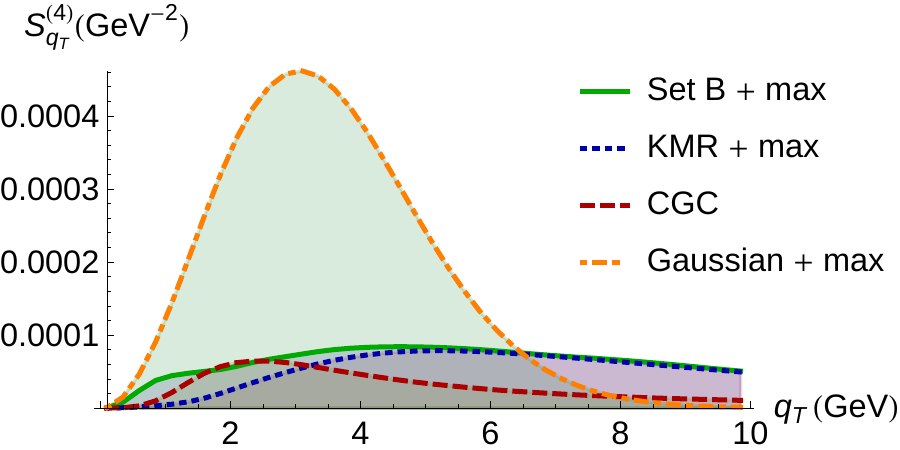}}
\caption{Model predictions for ${\cal S}^{(0)}_{q_T}$,  ${\cal S}^{(2)}_{q_T}$ and  ${\cal S}^{(4)}_{q_T}$ for the process $p(P_A)+p(P_B)\to {\cal Q}(P_{\cal Q}) + \gamma (P_{\gamma})+ X$  at $\sqrt{s}=14$ TeV in the kinematic region defined by  $Q=20$ GeV, $Y=0$, $\theta =\pi/2$,  and  $x_a=x_b \simeq1.4\times10^{-3}$.}
\label{fig:dsigma4dqT}
\end{figure*}

Our model predictions are presented in Fig.~\ref{fig:dsigma4dqT} for $\Upsilon+\gamma$ production, in a kinematic region where color octect contributions are suppressed~\cite{Qgamma}. The size of ${\cal S}^{(0)}_{q_T}$ should be sufficient 
to allow for a determination of the shape of $f_1^g$ as a function of $q_\sT$. Since ${\cal S}^{(2)}_{q_T}$ and ${\cal S}^{(4)}_{q_T}$ are considerably smaller, one would need to integrate them over $\bm q_\sT^2$  [up to $(Q/2)^2$] to get at least an experimental evidence of a nonzero $h_1^{\perp\,g}$.

\section{Conclusions}
The distribution of linearly polarized gluons inside an unpolarized proton $h_1^{\perp\,g}$ leads to a mo\-du\-la\-tion of the transverse momentum distribution
of scalar ($\chi_{c0}$, $\chi_{b0}$) and pseudoscalar ($\eta_c$, $\eta_b$) quarkonia that depends on their parity. It does not contribute to the transverse spectra of $\chi_{c2}$ and $\chi_{b2}$, which can be used to probe the unpolarized gluon distribution $f_1^g$. No angular analysis is needed for such measurements and experimental opportunities are offered by LHCb and the proposed fixed-target experiment AFTER at LHC~\cite{after}. Furthermore, a first determination of $h_1^{\perp \,g}$ and  $f_1^g$ could come from $J/\psi(\Upsilon) + \gamma$ production at the running experiments at the LHC, where yields are large enough to perform these analyses with existing data at $\sqrt{s}=7$ and $8$ TeV.  We have shown that, together with  similar studies in the Higgs sector, quarkonium production can be used to extract gluon TMDs and investigate their process and energy scale dependences. 

\section*{Acknowledgments}
This work was supported by the European Community under the “Ideas” program QWORK (contract 320389).

\end{document}